\documentclass{article}
\usepackage[cp1251]{inputenc}

\def\order#1.#2{\mathop{#1}\limits_{{\scriptscriptstyle #2}}\!\vphantom{#1}}

\def\frac#1#2{{#1 \over #2}}

\def\part#1#2{{\partial#1\over \partial#2}}

\usepackage{graphicx}

\begin{document}

\noindent{\Large Variations of the coefficient $J_2$ of
geopotential, and the dynamical Love number $k_2^d$
from the analysis of laser ranging to LAGEOS~1 and LAGEOS~2.
 }
\vskip2em

\centerline{\large G.A.Krasinsky} \centerline{\small\it Institute
of Applied Astronomy,
 St. Petersburg, Russia, email: kra@quasar.ipa.nw.ru}

\vskip4em

\newcommand{\beas}{\begin{eqnarray*}}
\newcommand{\eeas}{\end{eqnarray*}}
\newcommand{\bea}{\begin{eqnarray}}
\newcommand{\eea}{\end{eqnarray}}
\noindent

\noindent {\bf Abstract}
Secular and seasonal variations of the coefficient $J_2$ of the
geopotential are studied from the analysis of laser measurements
of distances to the geodetic satellites LAGEOS~1 (1988--2003) and
LAGEOS~2 (1992--2003). It is confirmed that beside the well-known annual
variations with the amplitude   $\approx 2.5 \times 10^{-10}$
there also exist very significant semi-annual variations of a comparable
amplitude.
 Phases of these two modes are such that
 the total effect may be described as a sharp postive splash of $J_2$
 in August  and considerably smaller variations in the rest part of year.
 The adopted  theoretical value of the so-called dynamical Love number $k_2^d$
 (a scale factor of
 near-diurnal oscillations aroused by the Earth's fluid core in the coefficients 
$c_2^1$, $s_2^1$ of the geopotential) is improved applying a simple close form 
of these oscillations expressed
 in terms of the differential angular velocities $v_1$, $v_2$ of the fluid core.
 It is shown that this form is equivalent to the standard one of Fourier series
 in which such oscillations
 usually are referred to as a frequency-dependence of the Love number $k_2$.
 The derived estimate $k_2^d=0.0595\pm 0.0007$ statistically differs from
 the theoretical value $k_2^d=0.063$. Out-phase oscillation of $k_2^d$ with the 
period about 18.6 years and the amplitude $0.0064 \pm 0.0011$ is detected giving 
evidence of large dissipation in the fluid core.
 The estimated secular trend $\dot J_2$ (commonly interpreted as the
 effect of the so-called Post-Glacial Rebound) appears twice less than the value
 recommended by the standards of the International Earth Rotation Service (IERS)
 but agrees with  last findings of other authors.

\vskip2em


\section{Introduction}

Starting point of this research was studying the near-diurnal oscillations 
$dc_2^1$, $ds_2^1$ caused by the Earth's fluid core in the coefficients 
$c_2^1$, $s_2^1$ of geopotential.
This effect  (interpreted as the frequency-dependence of the Love number $k_2$)
is presented by IERS standards in the form of the sum of trigonometric harmonics 
that depend on the fundamental arguments of the nutation theory  (McCarthy, 
Petit, 2004). It may be shown that the amplitudes of these harmonics are 
proportional to the so-called dynamical Love number $k_2^d$. In Appendix, the 
expression for the corrections $dc_2^1$, $ds_2^1$ is derived  in the simple 
close form:

\bea \pmatrix {d c_2^1 \cr
 d s_2^1 \cr }
=-\omega^{-1} J_2 {k_2^d\over k_s } \pmatrix { v_1
\cr v_2 \cr }, \label{v12} 
\eea
where $v_1$, $v_2$ are the
equatorial components of the angular velocity of the fluid core
relatively to the Earth as a whole, $ J_2$ is the coefficient of the
second zonal harmonics of the geopotential, $\omega$ is the Earth
angular velocity,  $k_s=0.93831$ is the so-called secular Love number. It will 
be also shown that this presentation is equivalent to the standard form of 
trigonometric series for the frequency-dependence of the Love number $k_2$.

The parameter $k_2^d$ plays important role in two geodynamical
problems. Firstly, it is rotation of the deformable Earth in the
framework of the standard SOS model (Moritz, Muller, 1987), and
secondly, dynamics of the Earth's satellites. In short,
perturbations of rotation of the deformable Earth depend on the
so-called compliance $\xi$ that describes the tidal response of
the moments of inertia of the Earth on the differential rotation
of its fluid core. This parameter may be expressed in terms of the
dynamical Love number $k_2^d$ by the relation $\xi=e k_2^d/k_s$ in
which $e$ is the dynamical flattening of the Earth. Unfortunately,
while processing Celestial Pole positions provided by
geodetic VLBI observations it proved  impossible to separate $k_2^d$
from the correction  to the ratio of the moment of inertia of  the
fluid core to that of the Earth as  a whole  (Shirai, Fukushima,
2001), (Krasinsky, Vasilyev, 2006).  On the other hand, the
near-diurnal variations  of the coefficients $c_2^1$, $s_2^1$ of the
geopotential bring about accumulating secular effects in  the node
of satellite orbits and thus one may hope to estimate $k_2^d$ from
satellite laser ranging (SLR) observations of geodetic  satellites.
As the adopted value of $k_2^d$ is purely theoretical, confronting
it with  experimental data is quite an actual problem. The value
derived in this paper from the analysis of  SLR  data of the
geodetic satellites LAGEOS~1
 and LAGEOS~2 is, to our knowledge, a first  experimental estimation
 of this parameter.

>From the same analysis of the SLR data, we determine also
corrections to the coefficient $J_2$ of the geopotential for studying
the time-variability of this coefficient. In such a study,
obtaining the secular trend $\dot J_2$ is a problem of primary
importance as the IERS standards  recommend to take this effect into
account whenever numerical integration of the equations of
artificial satellite motion has to be done. At present,  two centers of
analysis (University
of Texas and Goddard Space Flight Center) are monitoring the
geopotential on a regular basis. Recent publications of
the first group  (Cheng, Tapley, 2008) and the second one (Lemoine
et al., 2006) have revealed serious discrepancies  between the
derived values of the  secular trend: $\dot J_2=-2.6\times
10^{-11}\mbox{/year}$ and $\dot J_2=-1.3\times
10^{-11}\mbox{/year}$, correspondingly. So, the independent study
of $\dot J_2$ documented  in Section 3.3 seems to be actual.

\section{Observations and their processing}

We have used the laser observations of the geodetic satellites LAGEOS~1
(1988--2003), and LAGEOS~2 (1992--2003) taken from the server 
{\it ftp://cddis.nasa.gov}. The observations of each year were combined into one 
series and treated simultaneously. In Tab.1, the following information on the 
observations is given  for  each year: their total number $N_{obs}$, the total 
number $N_{del}$ of rejected observations, the Weighted Root Mean Square (WRMS) 
error of one-way distances to the satellite (in millimeters). Six orbital 
elements of the satellite under study and two dynamical empirical terms (along- 
and cross-track perturbing accelerations) were estimated weekly as local 
parameters. When processing the observations of each year, these local 
parameters were determined simultaneously with the global ones which are the 
coordinates of observing stations and the value of the dynamical Love number 
$k_2^d$. In addition, on each monthly interval (more exactly, on the interval 
of four weeks) a correction to the coefficient $J_2$ of the geopotential also 
was determined for further study of its time  variations. It is known that SLR 
observations made in different stations are of quite various quality. Properly  
down-weighted or rejected, the observations of poor quality do not distort the 
resulting estimates. In fact, only few stations of the last generation provide 
the overwhelming volume of high precision data that affect the results.

As an example, from the typical set of the observations of
LAGEOS~1 of the year 2000,
 the following parameters have been derived:

\begin{itemize}
\itemsep -1mm
\item
 dynamical Love number $k_2^d$,
\item
monthly corrections to $J_2$    (13 unknowns),
\item
 weekly corrections to six elements of the satellite and two
 empirical
parameters ($52 \times 8=416$ unknowns),
\item
yearly corrections to the coordinates of 99 stations (295
unknowns).
\end{itemize}

Thus, the total number of the estimated parameters for this year is 626.
Values of the longitudes and latitudes of several fiducial stations (which
provide the bulk of accurate observations) have been fixed to
prevent degeneration of the normal matrix due to correlations with
the elements of orientation of the satellite orbit. The software
complex ERA for ephemeris and dynamical astronomy was applied while
constructing the numerical theories of the orbital motion,
calculating the theoretical distances to the satellites and
processing the condition equations. The current DOS and Windows versions
are available
from the anonymous FTP server {\it quasar.ipa.nw.ru/incoming/era}.
All calculations are carried out in accordance with the
recommendations of IERS 2003 conventions (McCarthy, Petit, 2004).
The only difference in the dynamical equations of the satellite
motion is making use of the close form (\ref{v12}) for the
corrections $dc_2^1$, $ds_2^1$ caused by the fluid core. We will
show that this approach is equivalent to that recommended by IERS
but is more convenient because  the functional dependence of the
perturbing forces on $k_2^d$ is presented in  explicit form.

\begin{table}
\caption {WRMS errors $\sigma$ of the post-fit residuals (mm)}
\begin{center}
\begin{tabular}{|l|rrc|rrc|}
\hline \noalign{\vskip 1mm}
        &              & LAGEOS 1   &             &  & LAGEOS 2 &\\
 Year & $N_{obs}$ & $ N_{del}$ & $\sigma(mm)$& $N_{obs}$ & $ N_{del}$ & WRMS $(mm)$\\
\noalign{\hrule}
1988     & 67414      &   3058  &   22  &      &    &    \\
1989     & 64030      &   3411  &   22 &      &    &   \\
1990     & 77951      &   2721  &   22 &      &    &    \\
1991     & 54981      &   3518  &   22&      &    &   \\
1992     & 59715      &   6073  &   24  &  13683     &  863 &  21    \\
1993     & 81398      &   7665  &   22  & 78873     & 7128 &  22\\
1994     & 65370      &   4990  &   27 &  66827    &  4557 &  21   \\
1995     & 53463      &   4353  &   24  &  52153    &  3183 & 20   \\
1996     & 52725      &   4196  &   23  &  52244    &  7904  & 23   \\
1997     & 50404      &   3336  &   22  &  53516    &  3688  & 17    \\
1998     & 65010     &   5361  &   19   &  60819    &  3298  & 17  \\
1999     & 71788      &   4693  &   23  &  66520   &   3415&   18     \\
2000     & 61888      &   4122  &  23  &  63158    &  2362 &   19     \\
2001      & 71839      &   4060  & 19  &  70449    &  3083 &   19  \\
2002     & 70780      &   2503  &  21 &  64011    &  2217 &    18  \\
2003    & 74969      &   2379  &   19  &  74613    &      2798 &   19     \\

\noalign{\hrule}
 \end{tabular}
 \end{center}
\end{table}

\section{Results of data processing}

\subsection{Dynamical Love number $k_2^d$}

 Applying the method described above, we have obtained the set of 16 yearly 
estimates of $k_2^d$  from the observations of LAGEOS~1 (1988--2003) and 12 
estimates from LAGEOS~2 (1992--2003). They are presented in Fig.1
 (the black circles correspond to LAGEOS~1, the hollow ones to LAGEOS~2).
 Generally, all the estimates are in a very good accordance (the only exception 
is the LAGEOS~1 result for the year 1995 which evidently drops out and thus has 
not been used). From this set, the following averaged value of the dynamical 
Love number is derived

\beas
 k_2^d =
0.0595\pm 0.0007
 \eeas
that statistically differs from the  theoretical {\it a priory} value
$k_2^d=0.063$ given in (Moritz, Muller, 1987).

Fig.1 demonstrates also an out-phase oscillation with the period  18.6 years
and the amplitude $0.0064\pm 0.0011$. Probably, that is
 evidence
of large dissipation  in the fluid core.  Note that  parameters of the
dissipative effects in the fluid core  recommended by IERS standards are purely 
theoretical, and their numerical values have not yet been verified  by 
observations.

\begin{figure} 
\caption{Yearly estimates of dynamical Love number $k_2^d$}
\vskip 19.5mm
\hskip 10.0mm
\includegraphics[height=48mm]{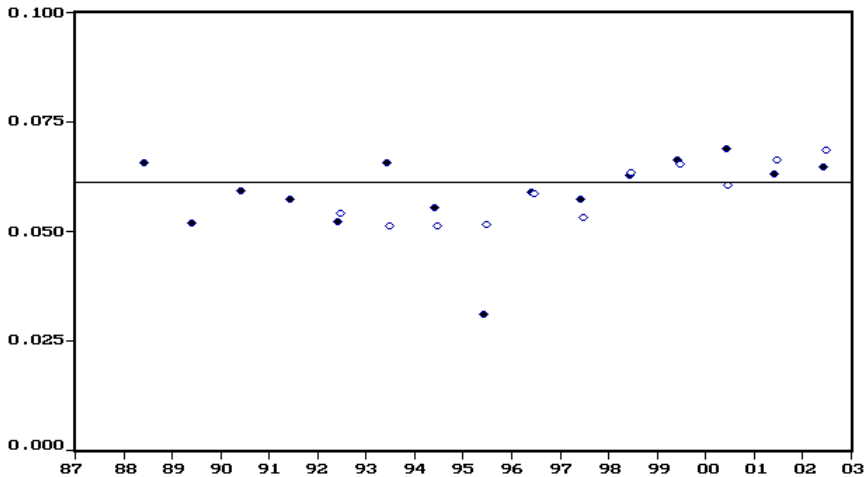}
\vskip 4.0mm
\end{figure}

For checking a possible dependence of $k_2^d$ on its starting value, the 
observations of an arbitrarily chosen year were processed with the zero starting 
value of $k_2^d$. After several iterations, the resulted estimate appeared the 
same.

 Note that in the methodically correct approach, the standard "static" Love 
number $k_2$ should also be estimated from the analysis of SLR data.
 At present, the most accurate and reliable estimate of $k_2$ might be obtained
 from the analysis of geodetic VLBI observations.
  Using these data,  the following consistent estimates have been derived:
 $k_2=0.2788 \pm 0.0011$ (Shirai, Fukushima, 2001) and $k_2=0.27272
 \pm 0.00036$ (Krasinsky, Vasilyev, 2006) which  are significantly less
 than the adopted value $k_2=0.3$. The error of this value may distort
 results of processing the SLR data.

\subsection{Periodic variations of the coefficient $J_2$ of geopotential}

Fig.2 presents the monthly corrections to $J_2$. The noticeable positive trends 
means that the {\it a priory} value  $\dot J_2=-26 \times 10^{-12}/\mbox{year}$ 
used in our calculations should be considerably diminished. When comparing Fig.2 
with the analogous plots in (Cheng, Tapley, 2004) and (Cox et al, 2005) it is 
necessary to bear in mind that Fig.2 presents corrections to the assumed IERS 
value of $\dot J_2$ while those in the cited works correspond to the zero value 
of it. More details concerning the value $\dot J_2$ are discussed in the next 
section.

 Methodically correct estimates of $\dot J_2$ may be derived  only 
simultaneously with determination of cosine and sine coefficients of the main
harmonics of the trigonometric polynomial approximation of $J_2$. These are the 
annual harmonics $A_{\cos}$, $A_{\sin}$, the semi-annual harmonics 
$A_{\cos}^{(0.5)}$, $A_{\sin}^{(0.5)}$, the harmonics $A_{\cos}^\Omega$, 
$A_{\sin}^\Omega$ connected with the period 18.6 year of revolution of the lunar 
node $\Omega$, and the empirical harmonics $A_{\cos}^{(3)}$, $A_{\sin}^{(3)}$ 
with  the period of 3 year. We also estimated the correction $dJ_2(0)$ to the 
value $J_2$ for 2000.0. Four solutions are obtained for different sets of the 
estimated parameters (see Tab.2).
 The first column of Tab.2
 specifies  parameters under estimation,  others give
 their estimated values. For the secular trend, there are given both the
 correction $d\dot J_2$ to the {\it a priory} value
$-26 \times 10^{-12 }\mbox{/year}$, and the  full value $\dot J_2$.

Tab.2 demonstrates that the estimates of the annual
 and semi-annual harmonics are robust and only slightly depend on the solutions.  
For the annual oscillations, the results can be compared with those  by other 
authors. The full amplitude of the annual oscillations is $(229\pm 23)\times 
10^{-12}$ in accordance with the value $300\times 10^{-12}$ taken approximately 
from Fig.1 of (Cox et al., 2005). Similar results are obtained in a number of 
other works; for instance, in (Cheng, Tapley, 2004) the seasonal amplitude is 
$290 \times 10^{-12}$ (no error is given). The phases of the oscillations also 
agree: the negative minimum in January and the positive maximum in July.
 The  amplitudes of the semi-annual oscillations are rather significant and 
comparable with those of the annual ones.
The derived amplitudes of the semi-annual variations   are consistent
with the results published by other authors. For instance, Table 1
of paper (Nerem et al, 2006.) gives the estimated cosine and sine
 amplitudes
 of the semi-annual variations
 as $-2.41\times 10^{-10}$ and $0.78 \times 10^{-10}$ in 
accordance with our  findings (see Table 2).

  The solid curve in Fig.2 presents Solution~4
 (with maximal number of estimated parameters).  Fig.2 shows that the seasonal 
variations cannot be described as a simple harmonic oscillation but have more 
complex structure. Namely, they keep near constant value from the beginning of 
every year till August when a sharp maximum takes place, and then decrease up to 
the end of the year. The plot given by Fig.2 is very similar to the intra-annual 
variations of $J_2$ presented by Fig 3(a) of (Moor et al. 2006) and derived from 
Lageos 1,2 data for the time span 1998-2004.
 Such behavior is a consequence of the large value of the semi-annual amplitude. 
This feature of the intra-annual variations qualitatively agrees with Fig.3 of 
paper (Lemoine et al. 2006) where they are presented for the interval 2000--2007 
being derived from SLR data for 9 satellites combined with DORIS data. Note that 
Fig.3 in (Lemoine et al. 2006) presents variations of the coefficient 
$c_{20}=-J_2$. It is interesting that Fig.2 of the same paper does not 
demonstrate  the fine structure of the intra-annual variations, probably because 
time-intervals between  the consequent samples are two months (though the 
caption indicates one month, probably erroneously). With so low time resolution, 
the fine structure of the intra-annual variations cannot be discerned.
Supposedly, this fine structure is due to the seasonal asymmetry of ice cover 
developing in the southern and northern semi-spheres.
 For a more detailed study of  these variations, Fig.3 presents them vs. JD 
(mod 365.25) after removing the secular and long-periodic variations taken from 
Tab.2. Beside the maximum in August, this plot demonstrates also and l
ong-periodic variations taken from Tab.2. Beside the maximum in August, this 
plot demonstrates also two small minima at April and November.
 Here only the LAGEOS~2 results have been used as the more accurate ones. The 
bars present  statistical errors obtained after averaging 13 groups of the 
monthly estimates  $J_2$ for all the years.

In Solutions~3 and 4, the sine and cosine components of the oscillation with the 
period 18.6  years have also been estimated (with  the resulted WRMS errors 
considerably decreased). The nature of the empirical oscillation of the 3-year 
period estimated in Solution~4 is unclear. We tried several empirical 
long-periodic harmonics but statistically significant amplitudes were found only 
for the 3-year period.

 Fig.2 demonstrates that the  positive outliers correspond to the August spikes. 
It means that  the values of these sharp spikes vary with time and cannot be 
presented by the simple model as the sum of the annual and semi-annual 
oscillations. Probably, this effect should admit some geophysical 
interpretation.

In the last two lines of Tab.2, the WRMS errors of the residuals are presented 
(both for LAGEOS~1 and LAGEOS~2). Note that the WRMS errors for LAGEOS~2 are 
considerably smaller than those for LAGEOS~1. Formal erors of $J_2$ appeared to 
be by order less than the range of the fluctuations of $dJ_2$, and so could not  
be presented by error bars in Fig.2.

\begin{figure} 
\caption{Corrections to the nominal $J_2$ (for the adopted
 $\dot J_2$) in $10^{-10}$}
\vskip 19.5mm
\hskip 10.0mm
\includegraphics[height=48mm]{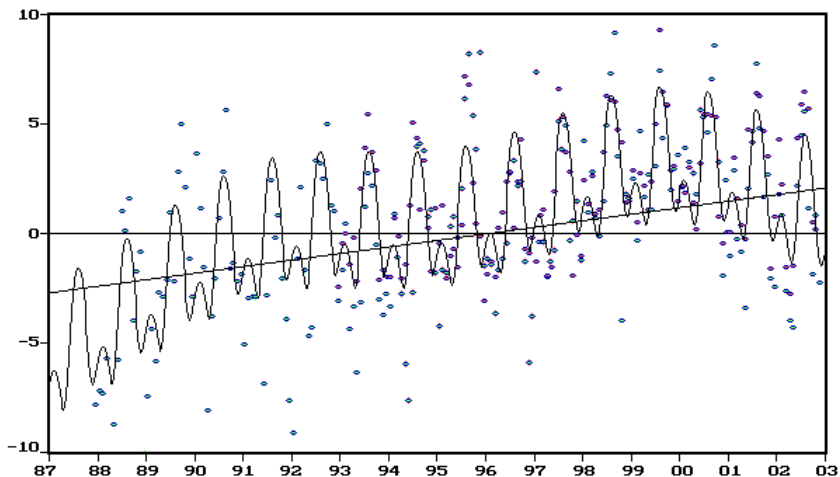}
\vskip 4.0mm
\end{figure}

\begin{figure} 
\caption{Intra-annual variations of $J_2$ vs. JD (mod 365.25) (in
$10^{-11}$)}
\vskip 19.5mm
\hskip 10.0mm
\includegraphics[height=48mm]{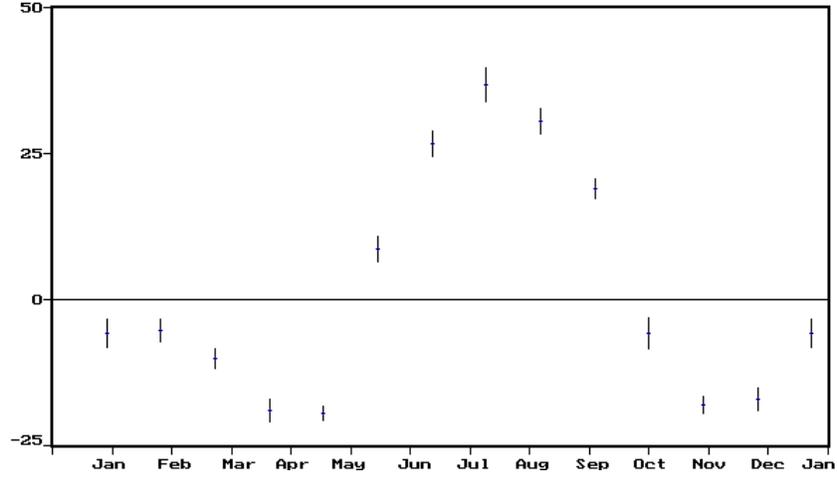}
\vskip 4.0mm
\end{figure}

\begin{table}
\caption {Models of variations of $J_2$,
  four solutions (in $10^{-12}$)}
\begin{center}
\begin{tabular}{|l|r|r|r|r|r|}
\noalign{\hrule} \hline \noalign{\vskip 1mm}
  & 1    &  2   & 3    & 4  \\
\noalign{\hrule} \hline \noalign{\vskip 1mm}
$dJ_2(0)$                   &    218(16) &      215(14) &   103(31) &  77(33) \\
$d\dot J_2/\mbox {year}$    &      33(3) &        35(3) &     18(7) &   13(7) \\
$\dot J_2/\mbox{year}$      &       7(3) &         9(3) &     -8(7) &  -13(7) \\
$A_{\cos}$                  &   -172(19) &     -174(17) &  -184(15) & -183(15) \\
$A_{\sin}$                  &   -151(19) &     -150(16) &  -164(15) & -168(15) \\
 $A_{\cos}^{(0.5)}$      &            &       66(16)  &   67(15)  &  66(15) \\
 $A_{\sin}^{(0.5)}$      &            &     167(17)  &  168(15)  & 168(15) \\
  $A_{\cos}^\Omega$          &           &              &  154(39)  &  181(40) \\
 $A_{\sin}^\Omega$          &            &               & -102(20) &  -98(20) \\
 $A_{\cos}^{(3)}$               &            &              &           &  -43(15)  \\
$A_{\sin}^{(3)}$                 &           &              &            &  -39(16) \\
\noalign{\hrule}
 WRMS, LAGEOS~1              &    285   &  255          &    245    &  235        \\
 WRMS, LAGEOS~2              &   195    & 155          &    135    &  135      \\
\noalign{\hrule}
 \end{tabular}
 \end{center}
\end{table}

\subsection{Secular trend $\dot J_2$ and the post-glacial rebound}

True value of the trend $\dot J_2$ is quite important as for practical applications as for a geophysical
interpretation. Indeed, the estimate $\dot J_2=-2.6 \times 10^{-11}/\mbox{year}$ is still recommended
by IERS standards (McCarthy, Petit, 2004) as experimentally confirmed effect of the post-glacial
rebound, despite that this trend has greatly diminished or even completely vanished for the SLR data
after 1998 (Cheng, Tapley, 2004). For instance,
Fig.3 in (Chao, 2003) shows that  the  accumulated error of $J_2$ becomes intolerably
large at present epoch if one applies the adopted value of $\dot J_2$.
 In recent publication  (Cheng, Tapley, 2008), the authors insist that this
value is still valid and explain  the clearly seen deviations of $J_2$ from observations after 1998
by some decadal variations, making stress on the effect of El~Ni\~no.  On the other hand,
the independent analysis carried out in Goddard Space Flight Center
(Lemoine et al, 2006) demonstrates  that  the time-behavior of $J_2$
was rather stable
 since 1976 up to 2007
with no noticeable change of it  after 1998. The resulting value
 $\dot J_2=-1.32 \times 10^{-11}$/year  is twice less than that of the IERS
standards based on the works by Cheng and Tapley.  Our analysis of
SLR data confirms this conclusion.  In Solution~4 (maximal
numbers of the estimated harmonics in $J_2$) we have obtained the estimate
$\dot J_2 =(-1.3 \pm 0.7) \times 10^{-11}$/year which is consistent with the result
of (Lemoine et al, 2006).
 Unfortunately,
due to the comparatively short 16-yearly time-interval of the data
used, the correlations of the amplitudes of the long-term
variations with the secular trend are rather large.
That is why the statistical error of our estimate of $\dot J_2$
is comparatively large.
In any case,
 we  may state that our value of $\dot J_2$ rules out
 the rate   $-2.6\times 10^{-11}/\mbox{year}$ recommended
 by IERS standards.

\section{Concluding notes}

1. From the practical point of view, the most important result of
this study is the conclusion that the derived
 secular trend $\dot J_2$ is consistent with the result
 of (Lemoine et al, 2006) confirming that the value of this trend recommended
 by IERS standards is twice too large.  To our opinion,  poor
 predictability  of the variations of $J_2$ due to
 inter-annual oscillations of unclear origin necessitates
  monitoring of the current values of $J_2$
  and disseminating the results for practical applications, just as at present
 it is done for the Earth's orientation parameters.

2.  The intra-annual variations of $J_2$ are not a
 simple harmonic oscillation; see Fig.3. Probably, the sharp spike  in August
 is due to asymmetric seasonal developing of the ice regime in the southern
 and northern semi-spheres. This problem deserves thorough study and a geophysical
 interpretation.

3. The derived estimate $k_2^d=0.0595 \pm 0.0007$ statistically differs from its theoretical
 value $k_2^d=0.063$. To our knowledge, that is the first experimental determination
 of this parameter.

 4. For independent calculation of various tidal effects caused by
 the fluid core, it is desirable to provide users by numerical values of
 the angular velocities $v_1, v_2$ of the fluid core.
 To avoid somewhat bulky analytical manipulations directly in the text of
present paper,
 derivation of the
 analytical expressions for  $v_1, v_2$ is transferred into Appendix.

\section{Appendix. Dynamical Love numbers}

It is known that the positions of sites on the Earth's surface are displaced by the pole tides of
 two types. Those of the first type are caused  by the motion of the pole of the  Earth as a whole,
 the corresponding tidal amplitudes being proportional to the Love numbers $h_2$, $l_2$.  The pole
 tides of the second type are caused by the differential rotation of the fluid core relatively to
 the mantle; their amplitudes are proportional to the so-called dynamical Love numbers $h_2^d$, $l_2^d$.
  The pole tides of the both types also contribute to the tesseral coefficients $c_2^1$, $s_2^1$ of
 the  geopotential. Putting aside the Chandler's free wobble of the pole, these contributions
   are of near-diurnal periods with the amplitudes proportional either
    to the static "potential" Love number $k_2$ or to the dynamical Love number $k_2^d$, the latter
     effect being the largest.
A theory of the dynamical Love numbers is given in monograph (Moritz and
Muller, 1987).

 As the fluid core brings about diurnal oscillations of the coefficients  $c_2^1$ and $s_2^1$
 of the geopotential, it  gives rise to rather significant long-term and secular perturbations in
 the elements of the observed satellites (mainly in the node) thus making it possible to improve
 the adopted theoretical value of the scaling factor $k_2^d$ of these perturbations. It is commonly
 assumed that the fluid core rotates relatively to the mantle with the angular velocity
 $\overline v= (v_1, v_2, 0$), the equatorial components $v_1, v_2$ being given by the theory of
 the Earth's rotation. We denote $\overline \omega=(\omega_1, \omega_2, \omega_3)$ the vector of
 the angular velocity of the Earth as a  whole. For calculating the potentials $dW_v$ induced by
 the centrifugal
acceleration of the differential rotation of the fluid core,  the velocity of any point $\overline r$
 in  these two domains within the Earth (the mantle and fluid core) may be presented either
 as $\overline \omega \times \overline r$ or as
 $(\overline \omega + \overline v) \times \overline r$, respectively. Then the
centrifugal acceleration $\overline W$ within the  mantle may be presented as
\beas
 \overline W =
-\overline \omega  \times
   \left(\overline \omega  \times \overline r \right)
   = -\overline \omega
   (\overline r, \overline \omega)
   + \overline r \omega^2,
\eeas
while in the fluid core the corresponding expression has
the form
 \beas
  \overline W  = -(\overline \omega + \overline v) \times
\left[(\overline \omega + \overline v) \times \overline r \right]
= -(\overline \omega + \overline v) \left[
  (\overline r, \overline \omega) + (\overline r, \overline v)
  \right] + \overline r |\overline \omega+ \overline v|^2
\eeas
with the standard notations $(\overline p,\overline q)$
 and $\overline p \times \overline q$ for the scalar
and vectorial products
of two vectors $\overline p$, $\overline q$.

The terms along the vector $\overline r$ in these expressions do
not deform the incompressible Earth and may be disregarded. Then
ignoring the second-order terms, we can set $\overline W =
\mbox{grad}\ W$, the potential $W$ at the right part being given
by the expressions

\bea W = \left\{
\begin{array}{l}
 -{1 \over 2}  (\overline r, \overline \omega)^2, \
 R_c<r<R,
   \\
 -{1 \over 2} (\overline r, \overline \omega)^2
-(v_1 x_1+v_2 x_2) x_3 \omega,\
 r<R_c,         \label{wr}
\end{array}
\right. \eea with the notations $R$, $R_c$ for the radii of the
Earth and its fluid core, respectively.

Adding  the spherically symmetric term ${3\over 2} r^2 \omega^2$
to the right part (that does not affect  the distribution of the
density within the incompressible Earth) and denoting $\cos S=
(\overline \rho, \overline \omega)/ \omega$, $\overline \rho=\overline r/r$,
 we can present the potential $W$ in each of these two domains within the Earth as
 the following combinations of the zonal and tesseral harmonic functions:

\beas W =
 \left\{
 \begin{array}{l}
 -{1 \over 3} \omega^2 r^2 P^0_2(\cos S),\   R_c<r<R,
  \\
 -{1 \over 3} \omega^2 r^2 P^0_2(\cos S)
-(v_1 x_1+v_2 x_2) x_3 \omega,\  R<R_c
\end{array}
\right. \eeas

In accordance with the general theory of Love numbers, action of
the perturbing spherical harmonics deforms the Earth's interior
and the resulting deformations induce the additional potential
$dW$ given on the Earth's spherical surface (of the mean radius
$R$) by the expression

\bea dW|{R} &=& -{k_2 \over 3} \omega^2 R^2 P^0_2(\cos S)-
 k_2^d R^2 (v_1 \rho_1+v_2 \rho_2) \rho_3 \omega, \label{dv} \eea
in which $k_2, k_2^d$ are the standard static and dynamic Love
numbers,  $\overline \rho= \overline r/r= (\rho_1, \rho_2, \rho_3)
$ is the unit vector to the probing point out of the Earth. In the
outer space, the first term at the right part generates the
additional potential $dW_r$

\bea dW_r = -{1 \over 3} k_2  \omega^2  { R^5 \over  r^3}
P^0_2(\cos S), \  r \ge  R \label{dwr} \eea which is the result of
the tides aroused by rotation of the Earth as a whole, while
the second term (proportional to $k_2^d$)  brings about the tidal
potential $dW_v$ caused by the differential rotation of the fluid
core. The potential $dW_v$ has the form of the second-order
tesseral harmonics:

\bea dW_v &=& -k_2^d  {R^5 \over  r^3}  (v_1 \rho_1+v_2 \rho_2)
 \rho_3 \omega,\ r > R.     \label{dwf}
\eea

For our aims, the zonal part of the  potential (\ref{dwr})  may be ignored  as it
 generates nothing but the permanent tidal component of $J_2$. Thus
retaining only the tesseral terms  (and keeping the old notation
$dW_r$) we obtain
 \bea dW_r &=& -k_2  {R^5 \over
r^3}  (\omega_1 \rho_1+\omega_2 \rho_2)
 \rho_3 \omega,\ r > R. \label{dw}
 \eea

Both these potentials, $dW_r$ and $dW_v$,  should be added to the geopotential $V$  which may
 be written in the standard form

\bea V= {G m \over  r}  \sum_{k, j} \left({ R \over r} \right)^k
P_k^j(\cos \delta)
 (c_k^j \cos j \lambda  +  s_k^j \sin j \lambda ), \label{geopot}
\eea
where $\delta, \lambda$  are the latitude and longitude of the probing point,
 $P_k^j$ are associated Legendre functions.
In order to present potentials (\ref{dwf}) and (\ref{dw}) in the similar form,
the combinations $\rho_1\rho_3, \rho_2\rho_3$ also should be expressed in terms of
 the associated Legendre functions:

\beas \rho_1 \rho_3&=& \cos \delta \sin \delta \cos \lambda = {1
\over 3}
 P_2^1(\cos \delta)
                  \cos \lambda, \\
\rho_2 \rho_3&=& \cos \delta \sin \delta \sin \lambda = {1 \over
3} P_2^1(\cos \delta) \sin \lambda, \eeas and then we obtain

\beas dW_v = - k_2^d \left({\omega R^2 \over 3}\right) \left({R
\over r}\right)^3 P_2^1(\cos \delta) (v_1 \cos \lambda +  v_2 \sin
\lambda). \eeas

Defining the so-called secular Love number $k_s$ in the standard way

\beas
k_s=3{Gm J_2 \over R^3 \omega^2},
 \eeas
 the expression for $dW_v$ may be re-written in the form

\bea dW_v = -  {Gm\over r} \left({R \over r} \right)^2
\left({k_2^d \over k_s}\right) J_2
 P_2^1(\cos \delta) (v_1 \cos \lambda + v_2 \sin
\lambda)/\omega. \label{dwv} \eea

Such a presentation of the potential $dW_v$ demonstrates that its contribution into
 geopotential (\ref{geopot}) has the form of corrections $dc_2^1$, $ds_2^1$ to
 the coefficients $c_2^1$, $s_2^1$. These corrections may be written in the simple form:

\bea \pmatrix {d c_2^1 \cr
 d s_2^1 \cr }
=-J_2 \omega^{-1} {k_2^d\over k_s } \pmatrix { v_1
\cr v_2 \cr }. \label{dcv1}\eea

The analogous expression is valid for the corresponding corrections $dc_2^1$, $ds_2^1$ proportional
 to $k_2$:

\bea \pmatrix {d c_2^1 \cr
 d s_2^1 \cr }
=-J_2 \omega^{-1} {k_2\over k_s } \pmatrix { \omega_1
 \cr \omega_2 \cr }. \label{dcr1}\eea

In this relation we only account for the components due to the
forced variations of $\omega_1$, $\omega_2$ (the so-called
Oppolzer's terms) omitting the free Chandler's oscillation which
is of much longer period. Hereinafter, the components of the
near-diurnal oscillations in $c_2^1$, $s_2^1$ proportional to
$k_2$ will be referred to as the Oppolzer's tides.

One can see that the variables  $v_1, v_2$ and $\omega_1,
\omega_2$ in the right parts of relations (\ref{dcv1}) and
(\ref{dcr1}) indeed oscillate with near-diurnal periods. For
instance, the angular velocities $\omega_1, \omega_2$ in
expression (\ref{dcr1}) may be presented in terms of the sidereal
time~$s$ and the time derivatives of the slowly changing angles of
precession $\phi$ and nutation $\theta$ by the Euler's kinematic
relations:

\bea  \omega_1 &=& \phantom{-} \dot \theta \cos s +\dot \phi \sin \theta \sin s, \nonumber \\
       \omega_2 &=& - \dot \theta
       \sin s +\dot \phi \sin \theta \cos s. \label{dcr12}
\eea

 In analogous way, the components $v_1, v_2$ of the angular
velocity of the fluid core may be transformed from the Earth-fixed
frame into its components $n_1$, $n_2$ in the inertial frame:

\bea  v_1 &=& \phantom{-} n_1 \cos s + n_2 \sin s \nonumber \\
       v_2 &=& - n_1
       \sin s + n_2 \cos s. \label{dcv12}
\eea

Slowly changing variables $n_1, n_2$ can be expressed in terms of
the nutational coefficients (and of the precession rate)  making use
of the differential equations of rotation of the
deformable Earth with the fluid core; see (Moritz and Muller,
1987).

If dissipation in the fluid core is taken into consideration,
the variables $v_1, v_2$ should be  calculated for the delayed time $t-\tau$.
With all necessary accuracy, the linearized relations
$v_1(t-\tau)= v_1(t)-\tau \dot v_1$ and $v_2(t-\tau)= v_1(t)-\tau \dot v_2$
may be applied. Moreover, to calculate the time-delayed variables $v_1$, $v_2$,
one can take into account only the time delay in the sidereal time~$s$
replacing in expression~(\ref{dcv1})  $v_1$ by $v_1-v_2 \delta$   and
$v_2$ by $v_2+v_1 \delta$ ($\delta=\tau \omega$ is the tidal phase lag).
The dissipation brings about out-of-phase terms in the trigonometric presentation
of the tidal effects. In the IERS standards, they are given for a theoretical value
of the phase lag  which, unfortunately, has not yet been verified
by satellite data.

Perturbing potential
(\ref{wr}) of the centrifugal accelerations brings about not only the
near-diurnal tidal variations (\ref{dcv1}), (\ref{dcr1}) of the tesseral coefficients
$c_2^1$, $s_2^1$ of the geopotential, but also near-diurnal displacements of sites: $dH$, $dN$, $dE$
 in the radial, northern and eastern  directions, respectively. In our notations, the displacements
 caused by the fluid core may be written in the form

\bea dH &=&   -R\left({h_2^d \over k_s}\right) J_2  \sin 2 \delta
(v_1 \cos \lambda + v_2 \sin \lambda)/\omega,
   \nonumber \\
dN &=&  -R\left({l_2^d \over k_s}\right) J_2 \cos 2 \delta (v_1
\cos \lambda + v_1 \sin \lambda)/\omega,
    \label{dN} \\
dE &=&   -R\left({l_2^d\over k_s}\right) J_2
     \sin \delta (-v_1 \sin \lambda + v_2 \cos \lambda)/\omega, \nonumber
\eea where $\lambda$, $\delta$ are the longitude and latitude of
the site, $h_2^d$, $l_2^d$ are the dynamical Love numbers.

The tidal displacements due to the centrifugal potential of the Earth as a whole may be obtained
 replacing the dynamical Love numbers $h_2^d$, $l_2^d$ by the static Love numbers $h_2$, $l_2$
 and the angular velocities $v_1$, $v_2$ of the differential rotation of the fluid core by
 the absolute angular velocities
$\omega_1$, $\omega_2$.
Broken into the trigonometric series of the fundamental arguments, the corrections (\ref{dcv1}),  (\ref{dN}) are interpreted in the
 IERS standards (McCarthy, Petit, 2004) as the effect of the frequency-dependence of
 the Love numbers $k_2$ (or rather $k_{21}$), $h_2$ and $l_2$. It will be shown
 that the explicit analytical expressions of these trigonometric
 series  have the following simple form (the proof is given at the end of Appendix):

 \bea dH &=& -R\left({h_2^d \over k_s}\right)J_2 \sin 2
 \delta \sum_{\nu} (C_\nu \cos \lambda + S_\nu \sin \lambda), \nonumber\\
  dN     &=& -R\left({l_2^d \over k_s}\right) J_2 \cos 2
 \delta \sum_{\nu} (C_\nu \cos \lambda + S_\nu \sin \lambda), \label{dnr}\\
  dE     &=& -R\left({l_2^d \over k_s}\right) J_2 \sin \delta
  \sum_{\nu} (-C_\nu \sin \lambda + S_\nu \cos \lambda), \nonumber
 \eea
\bea \pmatrix {d c_2^1 \cr
 d s_2^1 \cr }
=-J_2 \omega^{-1}\left( {k_2^d\over k_s }\right) \sum_{\nu}
\pmatrix {
 C_\nu \cr
 S_\nu \label{dcds}   \cr
 },
 \eea
  where
  \bea
   S_\nu &=& Q_\nu^+ \sin(s+f_\nu)+ Q_\nu^-  \sin(s-f_\nu), \label{snu}\\
   C_\nu &=& Q_\nu^+ \cos(s+f_\nu)+ Q_\nu^- \cos(s-f_\nu)  \label{cnu}
   \eea
   for $\nu \ne 0$,
   \beas
   S_0   &=& Q_0 \sin(s) \\
   C_0   &=& Q_0 \cos(s),
 \eeas
$s$ is the sidereal time, $f_\nu$ is the nutation argument of the
frequency $\nu$ in the nutational
 series given by the expressions
\bea \theta &=&  \sum_{\nu \ne 0} \delta \theta_\nu \cos f_\nu,
             \label{theta} \\
\psi   &=&   \sum_{\nu \ne 0} \delta \psi_\nu \sin f_\nu,
               \label{psi}
\eea
 $\delta \theta_\nu, \delta \psi_\nu$
are coefficients  of the nutations in inclination and longitude,
and the coefficients $Q_\nu^+$, $Q_\nu^-$, $Q_0$ are expressed in terms
of $\delta \theta_\nu, \delta \psi_\nu$ by the relations

  \bea
    Q_\nu^+  &=&(\delta \theta_\nu - \sin \theta_0 \delta
    \psi_\nu) q_\nu, \label{qnplus}
   \\
    Q_\nu^-  &=& (\delta \theta_\nu + \sin
 \theta_0 \delta \psi_\nu) q_{-\nu}\label{qnmin} ,\\
    Q_0 &=& - { p \over \nu_c }\sin \theta_0 (1-\kappa),
  \eea
 with the notations
 \bea
   q_\nu &=&  { \nu \over  \nu_c  + \nu } (1-\kappa), \quad \nu \ne 0,  \label{q_}   \\
   \kappa &=& \left({k_2^d \over k_s}\right) \alpha^{-1}, \label{kappa}
    \eea
 $p, \alpha, \nu_c$ being the precession constant, the ratio of the moment of inertia
 of the fluid core  to that of the Earth as a whole and  the Free Core Nutation frequency, respectively.

The corresponding effects of the Oppolzer's tides have the same
form and may be obtained  replacing $h_2^d$, $l_2^d$, $k_2^d$ by
$h_2$, $l_2$, $k_2$ and making use of the following expressions
for the coefficients $q_\nu$, $Q_0$:

 \beas q_\nu &=& \phantom{-} { \nu \over  \omega },  \quad \nu \ne 0,
       \\
 Q_0 &=& - { p \over \omega} \sin\theta_0.      \eeas

We prefer the close form (\ref{dcv1}), (\ref{dN}) of the
near-diurnal tidal effects instead of that recommended by IERS
standards because, firstly, it is much simpler, and,   secondly,
this formulation explicitly demonstrates that the variations of
$c_2^1$,    $s_2^1$ and the site displacements depend on the same
combinations of the variables $v_1$, $v_2$ (or $\omega_1$,
$\omega_2$) and differ only by constant factors. Another advantage
of this formalism written in the close form by relations
(\ref{dcv1}), (\ref{dcr1}), (\ref{dN}) is that it may be applied
not only with the analytical theories like IAU 2006 but also with
numerical theories of the precession-nutation motion like that
described in (Krasinsky, 2006), (Krasinsky, Vasyliev, 2006).
This approach guarantees the consistency of calculations, unlike
the formalism of IERS standards which suggests four different
trigonometric series for four tidal effects under consideration.
It is not clear how the recommended form of the near-diurnal tides
(in which the Oppolzer's tides are not separated from the tides
proportional to the dynamical Love numbers) may be applied
varying numerically the dynamical Love numbers in order to calculate the
corresponding partial derivatives.

It is easy to see that the numerical values of the diurnal tides presented in the form of
trigonometric series (\ref{dnr})--(\ref{kappa}) practically coincide with those given in
the IERS standards (McCarthy, Petit, 2004). As an example, in Tab.1 we
give the amplitudes $R_\nu$ (in mm) of the near-diurnal radial displacements calculated with
the above expressions as well as the corresponding numerical values taken from IERS standards, see
Tab. 7.5a of IERS Technical Notes 32 (McCarthy, Petit, 200).
These displacements depend on the dynamical Love number $h_2^d$ by relations

\beas \pmatrix {R_\nu^+ \cr
 R_\nu^- \cr } =-R
J_2{h_2^d\over k_s } \pmatrix { Q_\nu^+ \cr Q_\nu^-
\cr }, \eeas
where $Q_\nu^+,  Q_\nu^-$ are given by expressions (\ref{snu})--(\ref{kappa}),
 and the value of $h_2^d$ has been taken in such a way that the largest amplitude
(of the tidal constituent $\mbox{K}_1$ for which $\nu$=0) in the IERS standard would be equal
 to the sum of the corresponding terms proportional to $h_2^d$ and $h_2$. Tab.1 presents
 the coefficients $R_\nu$ giving not the frequencies $\nu$ but by the corresponding periods
 (providing also the integer coefficients in the linear combinations of the fundamental arguments
 $l$, $l'$, $F$, $D$, $\Omega$).  As the periods are given with their signs, the upper index
 in the coefficients $R_\nu^+$ or $R_\nu^-$ has been omitted. One can see that in the domain of
 the positive high frequencies $\nu$, the contribution of the Oppolzer's tides
 (the column $R_\nu^{(Opp)}$) is comparable with the effects of the fluid core. Tab.1 demonstrates
 that the IERS coefficients indeed are the sums of the two components: those proportional to $k_2^d$
 and to $k_2$, the first component being the largest.
 (The coefficients of IERS and ours in the two last lines slightly  differ).
Analogous comparison with IERS standards were carried out also for
the corrections $dc_2^1$, $ds_2^1$. The corresponding Fourier
coefficients differ only by a constant factor from
those of Tab.4 and  fixing properly this constant,
a good agreement has been reached again. However, now we are
interested in the absolute values  of the coefficients which
depend on the corresponding dynamical Love numbers. Serious
discrepancies were found in this case. Indeed, for the maximal coefficient
(the constituent $\mbox{K}_1$, the zero frequency $\nu$)  calculated with the
{\it a priory} $k_2^d$ we have obtained the value $589\times
10^{-12}$ which considerably exceeds the value $470\times
10^{-12}$ of IERS 2003. Making use of the new estimate
$k_2^d=0.0595$ derived in this work from the SLR data gives us the
value $546\times 10^{-12}$ of this coefficient which diminishes
the discrepancy but still significantly differs from the IERS
value. These considerations reinforce our conclusion on the
necessity of simultaneous estimating as $J_2$ as $k_2^d$.

\begin{table}
 \caption{Coefficients $\delta R_\nu$ of radial displacements (mm)}
\begin{center}
\begin{tabular}{rrrrrrrrr}
$R_\nu$ & $R_\nu^{(Opp)}$ & IERS    &  Period    &$l$  &$l'$  & $F$ & $D$ & $\Omega$   \\
              &    &  & days   &     &      &     &     &\\
  -0.04 & -0.04 &-0.08 &   9.13  & 1& 0& 2& 0& 2 \\
  -0.06 & -0.04 &-0.10 &  13.66  & 0& 0& 2& 0& 1  \\
  -0.28 & -0.22 &-0.51 &  13.70  & 0& 0& 2& 0& 2  \\
   0.04 &  0.02 & 0.06 &  27.60  & 1& 0& 0& 0& 0  \\
  -0.05 & -0.01 &-0.06 & 121.75  & 0&-1&-2& 2&-2  \\
  -1.20 & -0.10 &-1.23 & 182.62  & 0& 0& 2&-2& 2  \\
  -0.23 & -0.01 &-0.22 & 6798.38 & 0& 0& 0& 0& -1 \\
  11.72 &  0.28 &12.00 &   0.00  & 0& 0& 0& 0& 0  \\
   1.69 &  0.04 & 1.73 & -6798.38& 0& 0& 0& 0& 1  \\
  -0.70 &  0.00 &-0.50 & -365.30 & 0&-1& 0& 0& 0  \\
  -0.14 &  0.00 &-0.11 & -182.62 & 0& 0&-2& 2& -2  \\
   \end{tabular}
\end{center}
\end{table}

Now we derive in brief the expressions given above for the coefficients $C_\nu$, $S_\nu$ of
 the near-diurnal tidal effects. We will express the pole coordinates $v_1, v_2$ of the fluid core
 in terms of the coefficients of the adopted nutation $\delta \theta, \delta \psi$.  For the beginning,
 we restrict ourselves with the Poincare's model in which all tidal effects in $v_1$, $v_2$
 are disregarded. As the adopted nutation
theory has been constructed in the framework of a more realistic Earth's rotation model which
 does account for the perturbations caused by the non-zero Love numbers, in this way
 the indirect perturbations of this type are automatically accounted for. Though a more refine
 approach might be developed in which the explicit dependence of $v_1, v_2$ on the Love numbers
 would be also considered, only the main effect of this sort is taken into account in $Q_\nu^+$
 because
the resulting corrections are small enough. The following notations will be used:

\begin{enumerate}
\item
$ u=u_1+ i u_2$ and $v=v_1+iv_2$ are complex coordinates of the
pole of the Earth  and that of its fluid core,
\item $\omega$ is the angular velocity of the Earth,
\item $A, C$ are equatorial and polar moments of inertia.
\item $A_c, C_c$ are those of the core,
\item $e=(C-A)/A$, $e_c=(C_c-A_c)/A_c$ are dynamic
flatness of the Earth and that of its fluid core,
\item $\alpha=A_c/A$.
\end{enumerate}

In the Poincare's approximation, the differential equations
describing the time behavior of $u, v$
 may be written in the form
(Poincare,  1987):

\bea
 \dot u &=& \phantom{-} i u e  \omega -\alpha \dot v - i \alpha v + L \label{dotu0}\\
 \dot v &=& - i v (1+e_c) \omega  - \dot u,\label{dotv0}
\eea where $i=\sqrt{-1}$ and $L=L_1+i L_1 $ is a complex variable presenting the torques
caused by outer forces. Transforming these equations into the normal form in which
the time-derivatives  are in the left side, we obtain

\bea \dot u &=& \phantom{-} i u {  e \over 1-\alpha } \omega
   + i v {e_c  \alpha \over 1-\alpha} \omega  +{L \over 1-\alpha},
      \label{dotu} \\
\dot v &=& -i v \left( 1 +{e_c \over 1-\alpha} \right)\omega
  - i u { e \over 1-\alpha} \omega  - {L \over 1-\alpha}. \label{dotv}
\eea

Here the torque $L$  depends on the three Euler's angles: $\theta$
(the angle of nutation), $\psi$ (the angle of precession) and $\phi$ (the angle of the axial rotation).
 We can identify the variable $\phi $ with Greenwich Sidereal Time $s$.

Supplementing these equations with the Euler's kinematic relations
(\ref{dcr12}) rewritten in the form

\beas
u_1 &=& (\dot \psi \sin \theta \sin s  + \dot \theta \cos s) /\omega, \\
u_2 &=& (\dot \psi \sin \theta \cos s  - \dot \theta \sin s)
/\omega, \eeas we obtain a close system of differential
equations. The Euler's c relations  may be presented in the
equivalent complex form:

\bea
 u= D \exp(-i s), \label{utod}
\eea where \bea D=(\dot \theta + i \dot \psi \sin \theta )/\omega.
\label{D} \eea

The torque $L$ in equations (\ref{dotu}), (\ref{dotv}) depends on the Euler's angles in
 the following way:

\bea
 L= \exp(-i s) M(\theta, \psi) \label{m}
\eea where the function $M$ is independent of the sidereal time~$s$.

Defining new variable $Q$ by the relation
\bea v = \exp (-i s) Q
\label{vq} \eea
we can consider $D, Q $ as new variables and express differential equations (\ref{dotu}), (\ref{dotv})
 in terms of these variables. Calculating the time derivatives of the both sides of
 identity (\ref{utod}), we may ignore the time-dependence of
 the coefficient $D$. In the same approximation,
 $\dot s \approx \omega$ and we have

\beas
 \dot u &=&   (-i \omega D + \dot D) \exp(-i s) \approx - i
\omega D \exp(-i s). \eeas

With definition (\ref{m}), equations (\ref{dotu}), (\ref{dotv}) may be written now in
 the following form:

\beas
D &=&  - Q {\alpha e_c \over 1-\alpha}   + i {  M \over  \omega}, \\
\dot Q &=&  - i Q {e_c \omega \over 1-\alpha}  - i D
 {e \omega \over 1-\alpha} -   M.
\eeas

Eliminating the complex function $M$ from these equations, we obtain (neglecting the terms of
 the second order) the sought-for relation that ties the variables $Q$ and $D$

\bea
\dot Q = -i Q \nu_c  + i \omega D,      \label{dotQ}
\eea
with the notation $\nu_c$ for Free Core Nutation frequency:
\beas
\nu_c= e_c \omega.
\eeas

Let $\delta N= \delta \theta + i \delta \psi \sin \theta_0 $ be the complex nutation variable.
 Due to identity (\ref{D}), the variables $\delta N$ and $D$ are connected by the relation

\bea
D = {\delta \dot N \over \omega} - i { p \over \omega} \sin
\theta_0, \label{ddotn}
\eea
where the precession parameter $p>0$ is the coefficient of the linear trend in the precession
 angle $\psi$ (we use the right-hand coordinate frame in which the
precession motion is negative). The nutations $\delta \theta$, $\delta \psi $ may be presented
 in the form of the trigonometric series (\ref{theta}), (\ref{psi}).
In fact, the adopted nutations are referred to the instant coordinate frame,
 while we need the nutations in the fixed system of J2000. With the accuracy
 needed to calculate the small near-diurnal tidal effects, the difference may be ignored.
 The approximations $\dot f_\nu=\nu$ may be used due to the same reason. Now we present
 the complex nutational variable $\delta N$ in the form of the trigonometric series:

\bea \delta N= \sum_{\nu \ne 0} \left(
   N_\nu^+ \exp(i f_\nu) + N_\nu^-  \exp(-i f_\nu) \right),
\label{nsum} \eea where \bea N_\nu^+ &=&  {1\over 2} ( \delta
\theta_\nu + \sin \theta_0 \delta \psi),
\label{nsum+} \\
N_\nu^- &=&  {1\over 2} ( \delta \theta_\nu - \sin \theta_0 \delta
\psi). \label{nsum-}
\eea
The solution of equation (\ref{dotQ}) for $Q$ may be constructed in the similar form:

\bea
Q= \sum_{\nu \ne 0}  \left( Q_\nu^+ \exp(i f_\nu) + Q_\nu^-
\exp(-i f_\nu)
\right)
+ Q_0. \label{qsum}
\eea

Due to equations (\ref{ddotn})--(\ref{nsum-}), we can write the
coefficients $Q_\nu^+$, $Q_\nu^-$ and $Q_0$ as

\bea
 Q_\nu^+ &=&   {\nu \over \nu + \nu_c} N_\nu^+  \label{qunu+}, \\
 Q_\nu^- &=&   {-\nu \over -\nu + \nu_c} N_\nu^-,  \label{qunu-}
\eea
 \bea Q_0= -  \sin
 \theta_0 {p \over \nu_c }.
\label{nu0} \eea

Substituting relation (\ref {qunu+})--(\ref{nu0}) into (\ref{dcr1}), (\ref{dcv12}),
 we obtain the tidal effects under study in the form of relations (\ref{dcds})--(\ref{kappa}) for
 the Poincare's model (for which  the constant $\kappa$  defined by relation (\ref{kappa}) is
 equal to zero). The Poincare's model is the first approximation to the adopted model based on the so-called SOS
equations by Sasao, Okubo and Saito; see (Moritz and Muller, 1987). The next approximation
(which is sufficient for our aims) may be obtained by adding to the right part of equation (\ref{dotv0})
the term $\kappa L$. Repeating  the above transformations with the modified in this way
 equations (\ref{dotu0}), (\ref{dotv0}), we obtain the analytical expressions
(\ref{qnplus})--(\ref{kappa}) given earlier without proof.
In fact, for calculations of the present paper we have used not these analytical expressions
 but more accurate values $n_1$, $n_2$ of the angular velocity of the fluid core (in the inertial
 coordinate frame) obtained by numerical integration of the SOS-type differential equations described
 in (Krasinsky, 2006). The analytical expressions were used just for a control, in order to be sure that
 the calculated corrections are close to those given in IERS standards.

 \end{document}